# Teaching Fourier Analysis and Wave Physics with the Bass Guitar

Michael Courtney

Department of Chemistry and Physics, Western Carolina University

Norm Althausen

Lorain County Community College

This article describes a laboratory or demonstration technique employing the bass guitar and a Vernier LabPro (or a PC soundcard) for teaching wave physics and introducing Fourier analysis. The Fourier transform of an open string provides a demonstration of oscillatory modes out to the 20$^{th}$ harmonic consistent with expectations containing a fundamental frequency and harmonics. The playing of "harmonics" (suppressing resonant modes by lightly touching the string to enforce nodes at desired locations) demonstrates oscillations made up (mostly) of individual modes. Students see that the complete set of Fourier components (fundamental and harmonics) present on the open string can be explicitly connected with individual resonant frequencies as described in typical textbook discussions of natural frequencies of waves on a string. The use of a bass guitar rather than the six string electric guitar allows higher harmonics to be individually excited, and it is also easier for students to play the harmonics themselves.

0150Pa, 0230Nw, 0430Db

## Introduction

The Physics of standing waves on a string are described in many introductory textbooks. The wavelength, $\lambda_1$, of the fundamental mode of oscillation is twice the string length, and in general, the n$^{th}$ mode of oscillation has a wavelength $\lambda_n = \lambda_1/n$, where n is a positive integer. This produces resonant modes with frequencies $f_n = nf_1$, where $f_1$ is the fundamental oscillation frequency.

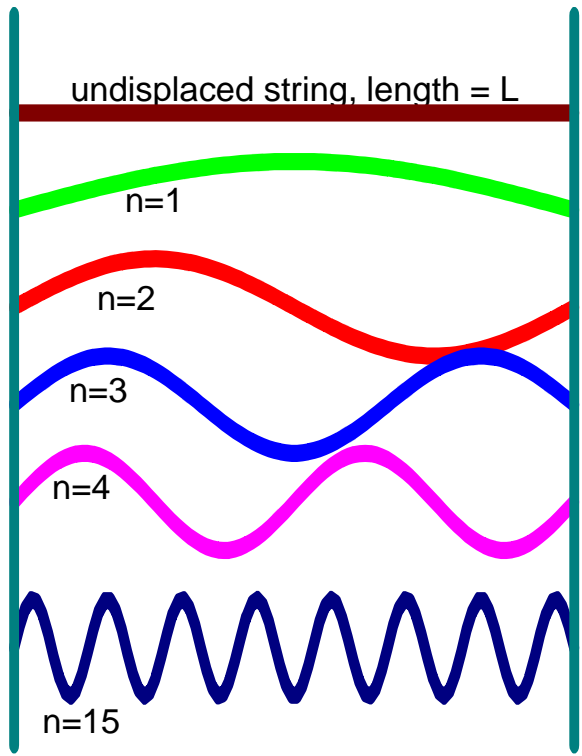

**Fig. 1: Shape of a vibrating string demonstrating a given resonant mode.**

A visualization of some resonant modes is shown in Fig. 1. In general, many resonant modes of oscillation are present in a vibrating string, and the resulting sound depends on the relative amplitudes and phases of the resonant modes. The bass guitar provides an example of waves on a string, and the resulting frequencies can be sampled simply with a microphone using either a Vernier LabPro and microphone and accompanying software, or a standard PC soundcard, microphone, and Audacity[1] software. Either approach allows for viewing of the Fourier transform of the sampled sound waveform representing the frequency spectrum of the sound that is produced from whatever oscillatory modes are present on the string. Therefore, the resulting frequency spectrum can be used to determine the presence of resonance modes on the string. This technique can be used to determine the frequency of resonant modes to within 2 Hz using a sampling time of 1 second.

**Resonant Modes in an Open String**

Playing the open A string results in the sound waveform and frequency spectrum shown in Fig. 2. The fundamental frequency is 55 Hz, and the first 20 harmonics are seen in the spectrum. When presenting this as a laboratory or a demonstration, it is helpful to have students predict the position of each successive peak before zooming in on the graph and verifying the expected frequency of the peak. One can also gear the demonstration to audiences less adept at math by tuning the A string a bit flat to 50 Hz so that the harmonic frequencies can be computed quickly by the audience.

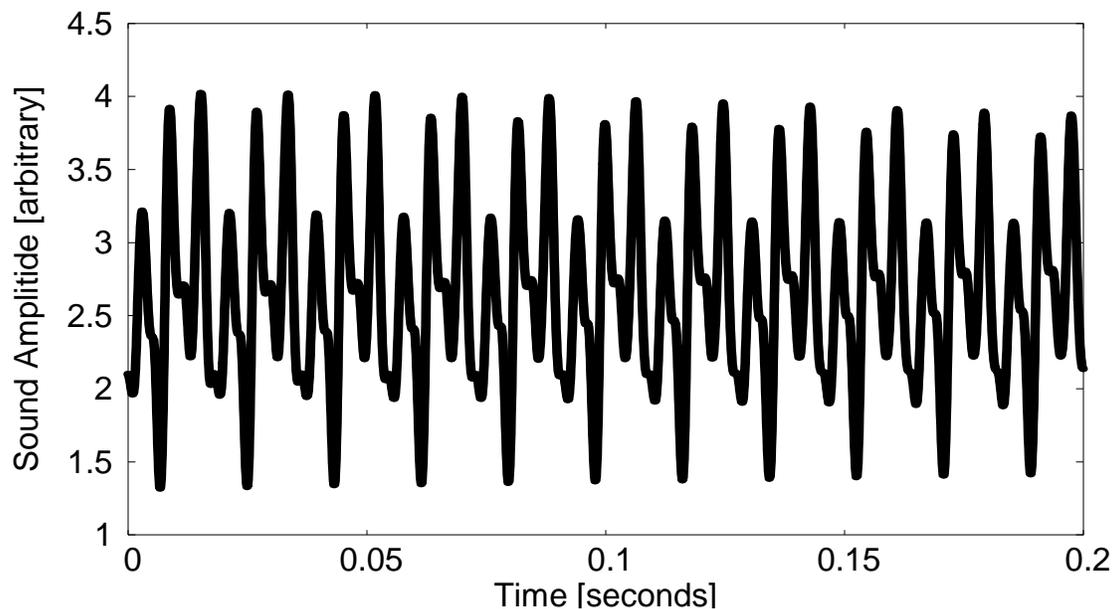

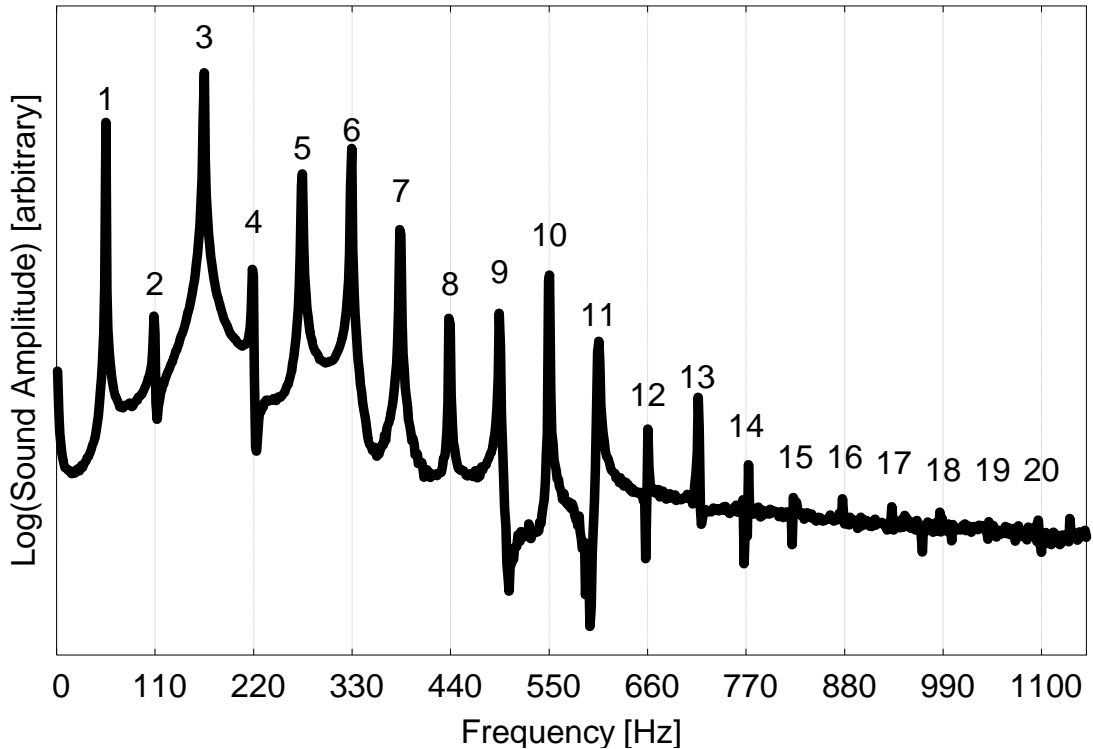

**Fig 2: Sound waveform and frequency spectrum of open string showing 1st 20 resonant modes of oscillation.**

Many introductory students are exercising faith in the instructor's expertise when making the connection that each peak in the frequency spectrum corresponds to a mode of oscillation such as pictured in Fig. 1. Playing "harmonics" on the bass guitar string by lightly touching the string at a given point to force a node at that point can be used to build a firmer association between the resonance mode shapes in Fig. 1 and the spectrum peaks in Fig. 2.

For example, lightly touching the string at its midpoint, *L/2* (just above the 12th fret) forces the string to have zero displacement at that point (a node). This removes all the odd harmonics from the spectrum and produces a spectrum with only even harmonics as shown in Fig. 3.

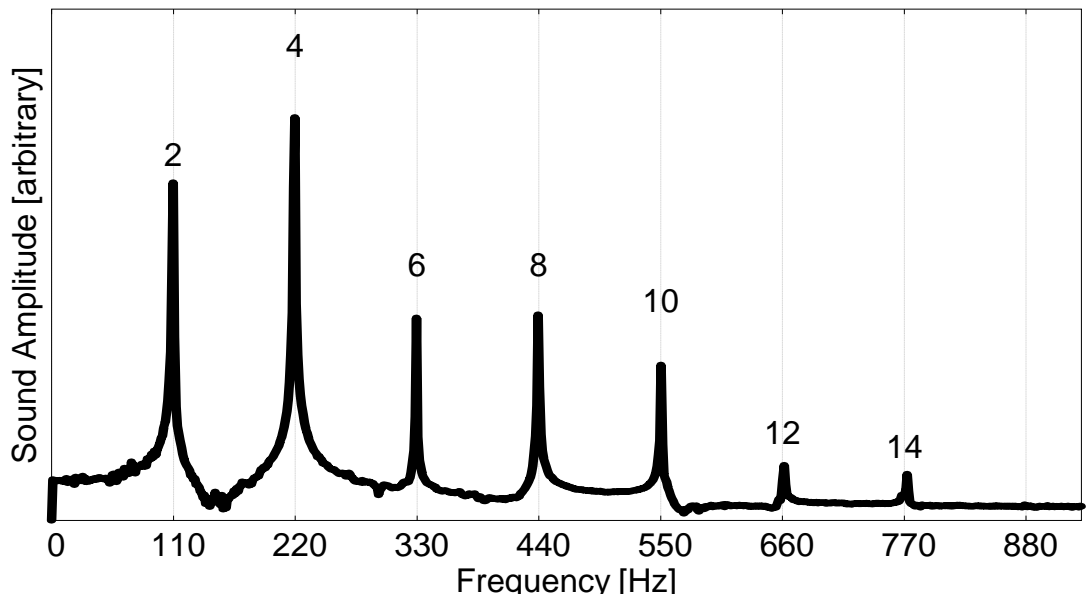

**Fig. 3: Frequency spectrum of open string with forced node at L/2 (12th fret) showing even modes of oscillation.**

This process of forcing a node at the point *L/n* can be repeated for *n=3, 4, 5* to suppress all the harmonics except multiples of n. Fig. 4 shows that forcing a node at *L/3* (just above the 7th fret) suppresses all the harmonics except multiples of 3. Likewise, forcing a node at *L/4* (just above the 5th fret) leaves only multiples of 4, and forcing a node at *L/5* (just above the 4th fret) leaves only multiples of 5. The correlation between the physical location of the node and the resulting spectra aids students in making the connections between the frequency spectrum and standing wave descriptions like Fig. 1.

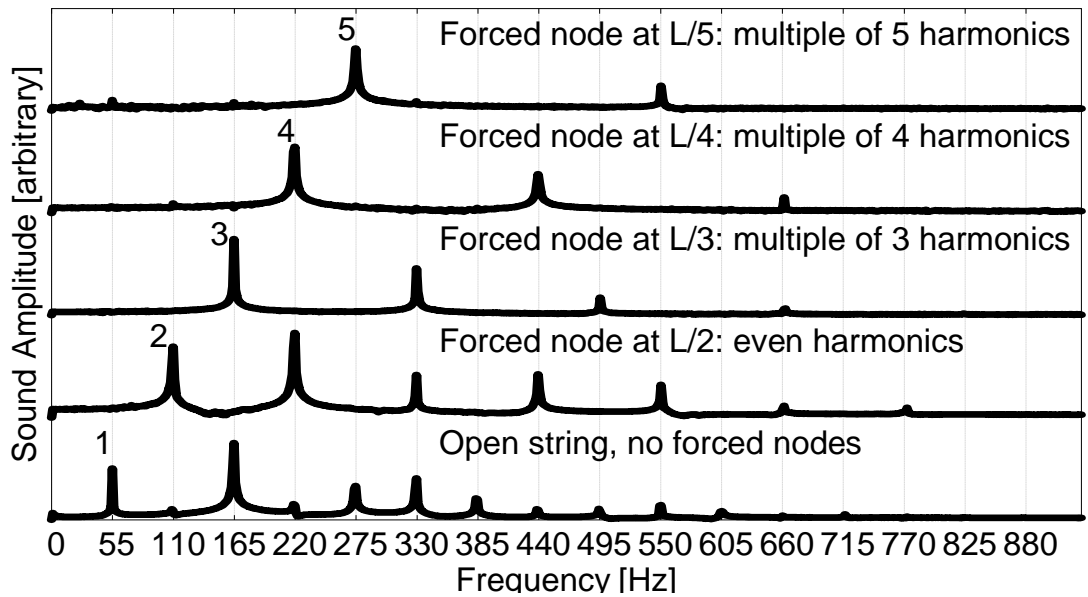

**Fig. 4: Frequency spectra comparison of varying locations of forced nodes.**

The process of forcing nodes at *L/n* can be repeated for n=6 through n=15 to demonstrate the $n^{th}$ harmonic in each case. Finding the correct location requires more careful measurement since the nodes are no longer located at frets. Successfully exciting single modes also requires progressively greater coordination between the plucking hand and the hand forcing the node. Fig. 5 shows that harmonics 6-15 can be excited relatively independently of other harmonics, although there is some energy in unintended modes, and the low-frequency background grows, particularly in the $14^{th}$ and $15^{th}$ modes.

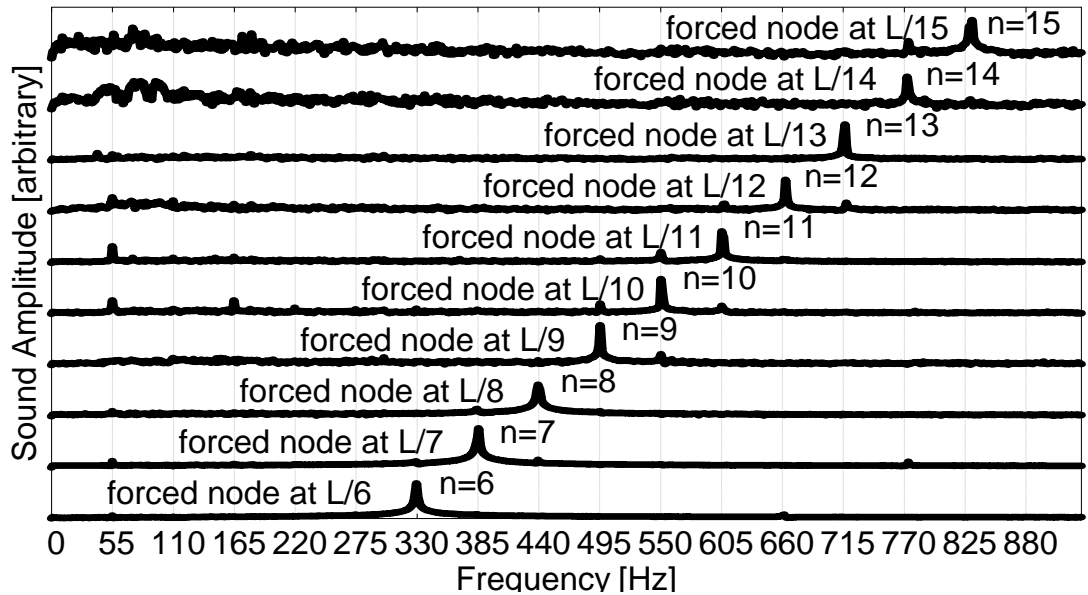

**Fig. 5: Forcing nodes at *L/n* for *n=6* through *n=15* suppresses resonant modes other than *n*, allowing spectrum to be dominated by mode at $nf_1$, where $f_1 = 55$ Hz.**

## Conclusion

The bass guitar is an excellent instrument for teaching wave physics and introducing Fourier analysis. When teaching this lab, we usually invite students to bring other musical instruments. This stimulates discussions of how various aspects of timbre, acoustics, and electronic effects pedals can be understood via Fourier analysis, as well as how electronic synthesizers produce a variety of realistic musical instrument sounds. A recorder or flute allows demonstration of standing waves in pipes. Having students speak or sing into the microphone demonstrates the basis of voice spectrogram identification. Most students agree that the musical instrument experiments make for the most exciting laboratories of the term.

The bass guitar is also an excellent vehicle for demonstrating important ideas for resonance energy transfer in photosynthesis and other phenomena.  When one plays an E note on any one of the three higher strings, the lowest string left open (also an E note) begins to vibrate also with an amplitude that is easy to detect.  In response to this demonstration, a student asked whether this effect is similar to a demonstration by his music teacher by pushing down a piano pedal and causing strings to vibrate by singing musical notes without touching the keys.  The light turns on.

---

[1] Audacity software available at http://audacity.sourceforge.net   This is a free audio capture, analysis, and editing program covered by GNU General Public License.